\documentclass[manuscript]{aastex}

\usepackage{natbib}
\usepackage{color}

\slugcomment{The Astrophysical Journal Letters}

\shorttitle{Multiple Slow Acoustic Oscillations in Stellar Loops}
\shortauthors{Srivastava et al.}

\begin{document}


\title{Evidence of Multiple Slow Acoustic Oscillations in the Stellar Flaring Loops of Proxima Centauri}


\author{A.K.~Srivastava\altaffilmark{1}}
\affil{1. Aryabhatta Research Institute of Observational Sciences (ARIES), Manora Peak, Nainital-263 002, India}

\author{S.~Lalitha\altaffilmark{2}}
\affil{2. Hamburger Sternwarte, University of Hamburg, Gojenbergsweg 112, 21029 Hamburg, Germany}

\author{J.C.~Pandey\altaffilmark{1}}
\affil{1. Aryabhatta Research Institute of Observational Sciences (ARIES), Manora Peak, Nainital-263 002, India}




\begin{abstract}
We present the first observational evidence of multiple slow acoustic oscillations in the post flaring loops of the corona of Proxima Centauri using XMM-Newton observations. We find the signature of periodic oscillations localized in the decay phase of the flare in its soft (0.3-10.0 keV) X-ray emissions. Using the standard wavelet tool, we find the multiple periodicities of 1261 s and 687 s. These bursty oscillations persist for durations of 90 minutes and 50 minutes, respectively, for more than 4 cycles.  The intensity oscillations with the period of 1261 s may be 
the signature of the fundamental mode of slow magnetoacoustic waves with the phase-speed of 119 km s$^{-1}$ in the loop of the length 7.5$\times$10$^{9}$ cm heated initially to obtain the flare peak temperature of 33 MK and later cooled down in the decay phase maintained at the average temperature of 7.2 MK.
The other period of 687 s may be associated with the first overtone of slow magnetoacoustic oscillations in the flaring loop. The fundamental mode oscillations show a dissipation with damping time of 47 min.
The period ratio P$_{1}$/P$_{2}$ is found to be 1.83 indicating that such oscillations
are most likely excited in longitudinal density stratified stellar loops.
We estimate the density scale height of stellar loop system as 22.6 Mm, which is smaller than the hydrostatic scale height of the hot loop system, and implies the existence of non-equilibrium conditions.
\end{abstract}



\section{Introduction}
Detection of magnetohydrodynamic (MHD) waves and oscillations may be very important in diagnosing the local plasma conditions of the 
coronae of Sun and Sun-like stars by impying the principle of MHD seismology 
(e.g., Nakariakov \& Verwichte 2005; Andries et al., 2009). MHD waves ({\it e.g.}, Alfv\'en, slow and fast magnetoacoustic waves) are well explored and studied in recent days in the solar atmosphere (i.e. in its chromosphere and corona) using the high resolution observations from Solar and Heliospheric Observatory (SoHO), Transition Region and Coronal Explorer (TRACE), Hinode, STEREO, and Solar Dynamics Observatory (SDO), along with the advancement  vis-a-vis theoretical modeling  and simulations (e.g., Aschwanden et al. 1999; Nakariakov et al. 1999; Ofman \& Wang 2002; O'Shea et al. 2007; Cirtain et al. 2007; De Pontieu et al. 2007; Verwichte et al., 2009, Jess et al. 2009; Srivastava \& Dwivedi 2010, Aschwanden \& Schrijver 2011; McIntosh et al. 2011, Kim et al. 2012; Mathioudakis et al. 2013 and references cited therein).  These waves can transport the solar photospheric powers into its chromosphere and corona in order to fulfill the fraction of energy losses.
The advanced  MHD seismology techniques have been elaborately developed to examine the crucial plasma conditions of the solar atmosphere based on the detection of the multiple harmonics of MHD waves (e.g., McEwan et al. 2006, 2008; Verth \& Erd\'elyi 2008; Srivastava et al. 2008, Andries et al. 2009; Wang 2011; Macnamara \& Roberts 2010, Luna-Cordozo et al. 2012; and references cited there).  

Complementing that the detection of MHD waves are getting enough importance in the development of the refined MHD seismology techniques to understand the properties of solar corona, it is also worth to state that there are several attempts to detect various MHD waves in the stellar coronae of Sun-like stars (e.g., Stepanov et al. 2001; Mathioudakis et al. 2003,2006; Mitra-Kraev et al. 2005; Pandey \& Srivastava 2009, Anfinogentov et al. 2013). The origin of such oscillations are still under debate, however, it is studied recently that they can be produced by several mechanisms during the flaring activities in the magnetized coronae of the Sun-like stars (e.g., Nakariakov \& Melnikov 2009, and references cited there). The widely accepted scenario behind the evolution of such quasi-periodic oscillations is the excitation of MHD wave modes in the stellar coronal structures. In the recent work, the slow mode oscillations are reported firstly by Mitra-Kraev et al.(2005) in the flaring loop of AT Mic, while the magnetoaocustic kink waves have been reported by Pandey \& Srivastava (2009) in the corona of $\xi$-Boo. Mathioudakis et al. (2006) have reported the observational scenario of very high frequency oscillations in the atmosphere of active star EQ Peg B, which may excite either due to fast MHD waves or due to the repetitive magnetic reconnection.  Recently, Anfinogentov et al. (2013) have found the evidence of various types of MHD oscillations in the dMe star YZ CMi.  However, most of the attempts in previous works were related to the detection of single MHD mode oscillations in the stellar coronae and their loops. To the best of our knowledge, there were no extensive efforts in the past to detect the multiple MHD modes and thus the based seismology of the stellar coronae.

In the present paper, we firstly detect the first two harmonics of the slow magnetoacoustic oscillations in the corona of a dMe star Proxima Centauri using observations carried out by XMM-Newton. 
The paper is organized as follows: in Sec.2, we present briefly the observations and adopted loop parameters. We describe the detection of MHD modes in Sec.~3. In Sec.~4, we explain the MHD seismology of the corona of Proxima Centauri.  In Sec.~5, the discussions and conclusions are outlined.

\section{Observations}

We choose a flaring epoch in the corona of Proxima Centauri observed with the XMM-Newton satellite using EPIC detector on March 13-14, 2009 (Obs ID - 0551120401).  Details of observations and data reduction are given in Fuhrmeister  et al. (2011).  The top panel of Fig.~1 shows the light curve of a flaring epoch of Proxima Centauri in the 0.3-10.0 keV energy band.
The flare starts at 06:00 UT, peaks at 06:45 UT, and show a long decay phase upto 09:00 UT on March, 14 2009.  The region between two dashed vertical lines is a region of our interest where we have searched for the localized MHD oscillations candidates. Bottom panel of Fig.~1 shows this localized temporal  span of the light curve during 12-21 ks in which the best fit exponential functions of the form $I=I_{o} e^{\pm(t-to)}/\tau_{d}$ are fitted to remove the long term background flare variations. Using the spectral modeling the flare peak  temperature was derived to be $33_{-11.93}^{+7.73}$ MK. However, average temperature during the post flare phase (i.e flare duration between 12-21 ks) was derived to $7.2\pm0.6$ MK (see Table 3 of Fuhrmeister  et al., 2011)
 
\subsection{Estimation of the loop length}
Using the time-dependent hydrodynamic loop model of Reale et al. (1997), Fuhrmeister et al. (2011) derived the loop half length of $8.55_{-2.86}^{+3.81}\times 10^9$ cm. This method includes both plasma cooling and the effect of heating during flare decay, however, sometime over-estimates the loop length (Reale 2007). If we assume that the flare emission originated from an ensemble of coronal loops with uniform cross section and roughly semicircular midplane shape, then we can express the observed  volume  emission measure (VEM) in terms of the loop length (L), density ($n_e$), number of loops (N), and aspect ratio ($\alpha$).  Following Huenemoerder et al. (2001), the loop length is derived as, L = $\pi^{-1/3}$$\alpha^{-2/3}$N$^{-2/3}$VEM$^{1/3}$$n_e^{-2/3}$. 
Using peak emission measure of the flare of $9.62\pm1.88\times10^{50}$ cm$^{_3}$ and density of $6.9\pm2.6\times10^{10}$ cm$^{-2}$ (Fuhrmeister et al. 2011), and assuming 100 flaring loops with aspect ratio of 0.1, the loop-length was calculated to be $4.0\pm1.0\times10^9$ cm. 
Using statistical loop length and peak temperature relation, $L = 10^{9.4} T_P^{0.91}$,  which leads the theoretically predicted scaling law  of $EM_P \propto T_P^{4.3}$ and further explains observed correlation of  $EM_P \propto T_P^{4.7\pm0.4}$ for both solar and stellar flares (Aschwanden et al. 2008), the loop length is calculated to be $7.5_{-2.2}^{+1.4} \times 10^9$ cm.  The loop lengths derived from above these two methods are consistent with each other. However, these loop lengths are  smaller than the one derived from a single-loop hydrodynamic model. Our derived loop length using the multi-loop model is though similar to those derived for other flaring dwarfs (Pandey \& Singh 2008), therefore for further analysis we have used the loop lenght of $7.5\times10^9$ cm.

\section{Detection of the Harmonic Periods in the Corona of Proxima Centauri}
We have used wavelet analysis IDL code $"$Randomlet$"$
that performs randomization test (O'Shea et al., 2001)
along with the standard wavelet analysis (Torrence \& Compo 1998) to examine
the statistically significant real periodicities in the time series data and to 
compute the associated powers. 
We use the flare light curve in 0.3-10 keV  energy band and 
vary time bin size from 50-90 s.
 We clearly notice 
a consistent and localized oscillatory pattern in the post flare phase of the emission (see Fig.~1).
We do not choose the lower binning as we were 
not interested in the very high-frequency oscillations of stellar loops. Using standard wavelet tool as discussed above, we 
derive the power spectra for each of these light curves during the post 
flare phase between 12-20 ks. We detect almost similar harmonic periods
in each binning of the data (cf., Table 1).
Here we present one example of the detected 
harmonic periods in Fig.~2 related with the 70 s binned data.

Depicted in Fig.~2 is the result from the wavelet analysis, in which the top 
panel displays the de-trended light curve of 70 s binned data during the post-flare phase 
heightened emissions (cf., top and bottom panels of Fig.~1; Sec.~1). 
In the middle-left panel  of Fig.~2, we plot the 
intensity wavelet that shows the bursty and localized nature 
of the oscillations and related powers. These oscillatory periods of 1261 \& 687 s were sustained for first 90 and 50 minutes with $>$ 4 cycles, respectively. The middle-right panel 
in Fig.~2 shows the global power spectrum with a significant periodicity of 1261 s with the probability grater than 99 \%.  We notice that the other bursty power lies at 687 s in the intensity wavelet and corresponding global power peak is also present. 
The bottom-panel displays the temporal variation of probabilities of the detection of these
periodicities also exhibit in localized region.
The other larger periods ($>$3000 s) may persist in the un-trended light curve, however, we searched for the periods that are naturally excited and associated with the possible MHD modes excited in the flaring stellar loops having certain morphological and plasma properties.
Therefore, the periods of our interest, which are 
most likely associated with the MHD activity of the loops, can be only detected after pre-whitening 
of these long term variations (O'Shea et al. 2007; Anfinogentov et al., 2013). 
The wavelet results are consistent with the Lomb-Scargle periodogram analyses of the same light curve (Scargle  1982).
In Fig.~3, we show the detection of almost similar periodicities near $\sim 1261$ s and $\sim 687$ s respectively that are above 1-$\sigma$ level. A higher period  of 8000 s is also seen in the power spectra which  is equivalent to  the total  data length. However, such long periods may not be the real and lie outside the COI in the wavelet.
Significant detection of naturally evolved periods in 50-90 s time-binned light curves
are summarized in Table 1. It should also be noted that the EPIC-MOS light curves of these
different binnings also exhibit the similar periodicities. 

\section{MHD Seismology in the Localized Corona of Proxima Centauri}
The observed harmonic periods (1261 and 687 s) may be  due to the impulsive generation of the slow MHD oscillations as a consequence of the flare energy release. 
The harmonic period of 687 s
is only present during first 3 ks duration (cf., Fig.2) 
which is likely to be an overtone that results from the initial impulsive excitation of the slow-mode wave, but damped out quickly than the fundamental period of 1261 s.
Observed oscillations may be excited in this loop due to the impulsive flare energy release and heating of the
loops (Nakariakov et al. 2004; Mendoza-Bricno \& Erd\'elyi 2006; Selwa et al. 2007). 
Using the loop length of $7.5\times10^9$ cm, phase  
speed of the fundamental mode MHD oscillations for a period of 1261 s is estimated to be 119 km s$^{-1}$, which  is sub-sonic in nature 
compared to the local sound speed (394 km s$^{-1}$) in the loop at average temperature of 7.2 MK during post-flare phase.
However, the peak flare temperature of 33 MK corresponds to the local sound speed of 
850 km s$^{-1}$. This impulsive heating may initially trigger the slow oscillations, however, they span
in the localized decay-phase where loop quickly cools down to be maintained at an average temperature
of 7.2 MK.
Therefore, the detected periodicities could  be due to the first two harmonics 
of the slow acoustic MHD oscillations in the coronal loops of Proxima Centauri.
In the case of slow magnetoacoustic oscillations density of the loop is perturbed 
due to the spatial redistribution of the plasma along the background magnetic field, i.e., 
the field-aligned movement of the plasma from one foot-point to the other in case of the fundamental
mode, and from both the foot-point to the apex in the case of its first overtone (Kim et al. 2012). Therefore, 
these MHD modes modulate the emissions of the plasma significantly. 
As the observed energy release is very powerful, the solar analogy suggests that it  
may also be a two-ribbon flare and the slow mode oscillations can be excited in a system of plasma arcades
(Nakariakov \& Zimovets 2011; Gruszecki \& Nakariakov 2011). However, the arcade model instead of 
the loop model does not change the above mentioned conclusions as the observations do not have spatial resolution.

Such type of the oscillations have already been detected in the hot and flaring loops
(Ofman \& Wang 2002, Nakariakov et al. 2004, Kim et al. 2012), as well as in the comparatively cooler non-flaring loops
(Srivastava \& Dwivedi 2010) in the solar corona. Such oscillations show 
the strong nature of the damping most likely due to the thermal conduction 
(Ofman \& Wang 2002). In the present case, the fundamental mode of the oscillations shows strong 
damping. In the case of the fundamental mode wave-period, the peak intensity decays upto
its 1/e value in $\sim$50 min, which is the damping time of such oscillations.

Fig.~4 shows the variation of exponentially decaying harmonic function along with the detrended light curve. The function is described as  


\begin{equation}
F(t)=\sum_{i=1}^{2} A_i cos \left(\frac{2\pi}{P_i}.t +\phi_i\right) e^{-\frac{\delta}{P_i}.t},
\end{equation}
where $A_{i}$ and $\phi_{i}$ are amplitudes and phases corresponding to the the oscillatory period, $P_{i}$ (1261s and 687 s), respectively, and $\delta$ is the damping factor. The best fit of the function (cf., Fig.~4) gives $\delta$=0.45, which indicates the damping time of 47 min.  It is well fitted the noisy light curve that used in the detection of these two periodicities (cf., Figs.~3-4). The  synthesized curve shows the damping period of 47 min, which is similar to the observed damping time of 50 min when the initial amplitude of oscillations decay to its 1/e value in the observational base-line (cf., top-panel of Fig.~2).
The observed damping time is almost in agreement with the scaling laws between 
the observed wave period and decay time of such oscillations as reported 
in the variety of hot loops in solar corona as observed by SUMER onboard SoHO (Ofman \& Wang 2002).
Therefore, we may also conclude that the strong damping of such oscillations
is due to the effect of thermal conduction in the loops of Proxima 
Centauri. 

We use the solar analogy 
to derive the plasma conditions of the corona of spatially un-resolved low mass flaring star Proxima Centauri (e.g., McEwan et al., 2006; Srivastava \& Dwivedi 2010; Macnamara \& Roberts 2010; Luna-Cardozo et al. 2012, and references cited therein).
The period ratio P$_{1}$/P$_{2}$ is derived to be 1.83. The period ratio less than 2.0 is most likely due to the longitudinal density stratification of the stellar loops of the star in which such oscillations are excited. The effect of density stratification on the lowering of the period ratio of first two harmonics of  slow acoustic oscillations has been only studied in the case of solar corona loops, which provides the most likely longitudinal density structuring of such loops (e.g., McEwan et al., 2006; Srivastava \& Dwivedi 2010; Macnamara \& Roberts 2010; Luna-Cardozo et al. 2012, and references cited there). 
It should be noted that lowering of the period ratio of first two harmonics
of slow waves only attributes to the density stratification. However,  
in the case of kink waves the period ratio can either be less than or greater than  2 implying that the density stratification and magnetic field divergence of the magnetic fluxtubes, respectively (Andries et al. 2009).

The present observations provide the first clues about the longitudinal density structuring in the stellar flaring loops. Therefore, such density or gravitational stratification should be considered in the future modelling 
of the stellar loops. Using equation (24) of McEwan et al. (2006), half loop length  of 3.75 $\times$10$^{9}$ cm, and period ratio (P$_{1}$/2P$_{2}$) of 0.915, the density scale height of the stellar loop system is estimated as $\sim$22.6 Mm. The density scale height is well below the hydrostatic scale
height of such loops assuming the typical coronal temperature of few mega Kelvin in Proxima Centauri during non-flaring state. This indicates the existence of non-equilibrium conditions there, e.g., flows and mass structuring, which is very obvious in the dynamical coronae of such stars. 
For example, the hot gas motions have been seen in the optical and
ultraviolet spectra of flaring  M dwarfs DENIS 104814.7-395606.1 and CN Leo (Fuhrmeister \& Schmitt 2004, Fuhrmeister et al. 2004). However, such physical scenarios are not well
established in the case of the stellar coronae (e.g., G{\"u}del \& Naz{\'e} 2010).


\section{Summary and Discussion}\label{SECT:DISS}
The heating due to the flaring activity either near the apex or near the footpoints 
of the loop can generate the various modes of the slow acoustic oscillations in magnetic loops (Nakariakov et al. 2004),
or in the loop-arcade systems in the flaring region (Nakariakov \& Zimovets 2011; Gruszecki \& Nakariakov 2011).
Random heating near the footpoint of these loops may also cause the excitation of the slow acoustic oscillations
(Mendoza-Brice{\~n}o \& Erd\'elyi 2006). 
Taroyan and Bradshaw (2008) have shown that the various harmonics of the slow-acoustic oscillations 
may be presented in the coronal loops maintained at the variety of temperatures, i.e., hot as well as cool loops. Later
the harmonic periods have been observed in cool non-flaring loops as observed by Srivastava \& Dwivedi (2010).
Previously, the slow acoustic oscillations were only discovered in the hot SUMER loops (Ofman \& Wang 2002).
However, the excitation conditions and typically the observations of such significant MHD modes are still under debate in the solar atmosphere. 

We present the first observational evidence of the harmonic periods (1261 and 687) of slow acoustic oscillations in flaring coronal loop system
of Proxima Centauri. The observed shorter periodicity (687 s) can be connected with the anharmonicity of the main periodicity (1261 s) and therefore
may not be associated with another mode of oscillations.
The appearance of such oscillations are bursty and temporally localized in the post flare phase that may be generated due to the 
impulsive heating of the stellar loops of proxima centauri after the flare energy release.
The fundamental mode oscillations show a strong decaying nature 
similar to the slow acoustic oscillations observed in the solar corona. Therefore, it may also be caused most likely due to
the thermal conduction (Ofman \& Wang 2002). One additional interesting property is evident in form of the period-shift of the 
fundamental mode slow acoustic  oscillations. The period is shifted towards higher values (cf., 
intensity wavelet of Fig.~2). This indicates the change in the local sound speed, which is the upper-bound of the phase speed of the 
detected slow magnetoacoustic oscillations in ideal plasma, as ${P[T(t)]}$$\propto 1/{C_{s}[T(t)]}$ (Aschwanden 2004). This means the 
local phase speed of slow acoustic oscillations will decrease with the shift of the period towards higher values upto
the time domain of 90 min. This indicates that the local ambient temperature of the loops may be decreased 
due to the radiative cooling and can cause the dissipation of these oscillations (e.g., Aschwanden 2004).
However, the radiative cooling may not have important role in dissipating 
the fast kink or sausage modes as typical time-scale of the 
radiative cooling (1 hr) in solar-type of loops may be higher compared to the 
damping time of these modes as theorized in various reports (Morton \& Erd\'elyi 2009;
Ruderman 2011).

This may be the most plausible non-ideal effect that can also cause the damping of such oscillations.
The period ratio P$_{1}$/P$_{2}$ is found to be 1.83, which
may be the most likely signature of density stratification in these stellar loops. The estimated density scale height is 22.6 Mm that is well below the hydrostatic scale
height of such loops. This
indicates the presence of non-equilibrium conditions in its corona, i.e, mass structuring as well as the flows.

\section{Acknowledgments}
We thank the  reviewer for his valuable suggestions  that improved the manuscript considerably, and Prof. Mihalis Mathioudakis for reading manuscript and providing useful suggestions. This work uses data obtained by XMM-Newton, an ESA science mission with instruments and contributions directly funded by ESA Member States and the USA (NASA).  AKS thanks Shobhna Srivastava for her support and encouragement during the work.

{}

\begin{thebibliography}{}
\bibitem[Andries et al.(2009)]{2009SSRv..149....3A} Andries, J., van Doorsselaere, T., Roberts, B., et al.\ 2009, \ssr, 149, 3 
\bibitem[Anfinogentov et al.(2013)]{2013ApJ...773..156A} Anfinogentov, S., Nakariakov, V.~M., Mathioudakis, M., Van Doorsselaere, T., \& Kowalski, A.~F.\ 2013, \apj, 773, 156 
\bibitem[Aschwanden et al.(1999)]{1999ApJ...520..880A} Aschwanden, M.~J., Fletcher, L., Schrijver, C.~J., \& Alexander, D.\ 1999, \apj, 520, 880 
\bibitem[Aschwanden(2004)]{2004psci.book.....A} Aschwanden, M.~J.\ 2004, Physics of the Solar Corona.
\bibitem[Aschwanden(2008)]{2008ApJ...672..659A} Aschwanden, M. J., Stern, R. A., G\"{u}del, M.\ 2008, \apj, 672, 659
\bibitem[Aschwanden \& Schrijver(2011)]{2011ApJ...736..102A} Aschwanden, M.~J., \& Schrijver, C.~J.\ 2011, \apj, 736, 102 
\bibitem[Cirtain et al.(2007)]{2007Sci...318.1580C} Cirtain, J.~W., Golub, L., Lundquist, L., et al.\ 2007, Science, 318, 1580 
\bibitem[De Pontieu et al.(2007)]{2007Sci...318.1574D} De Pontieu, B., McIntosh, S.~W., Carlsson, M., et al.\ 2007, Science, 318, 1574 
\bibitem[Fuhrmeister et al.(2004)]{2004A&A...417..701F} Fuhrmeister, B., Schmitt, J.~H.~M.~M., \& Wichmann, R.\ 2004, \aap, 417, 701
\bibitem[Fuhrmeister \& Schmitt(2004)]{2004A&A...420.1079F} Fuhrmeister, B., \& Schmitt, J.~H.~M.~M.\ 2004, \aap, 420, 1079 
\bibitem[Fuhrmeister et al.(2011)]{2011A&A...534A.133F} Fuhrmeister, B., Lalitha, S., Poppenhaeger, K., et al.\ 2011, \aap, 534, A133 
\bibitem[Gruszecki \& Nakariakov(2011)]{2011A&A...536A..68G} Gruszecki, M., \& Nakariakov, V.~M.\ 2011, \aap, 536, A68
\bibitem[G{\"u}del \& Naz{\'e}(2010)]{2010SSRv..157..211G} G{\"u}del, M., \& Naz{\'e}, Y.\ 2010, \ssr, 157
\bibitem[Huenemoerder et al.(2001)]{2001ApJ...559.1135H} Huenemoerder, D.~P., Canizares, C.~R., \& Schulz, N.~S.\ 2001, \apj, 559, 1135
\bibitem[Jess et al.(2009)]{2009Sci...323.1582J} Jess, D.~B., Mathioudakis, M., Erd{\'e}lyi, R., et al.\ 2009, Science, 323, 1582 
\bibitem[Kim et al.(2012)]{2012ApJ...756L..36K} Kim, S., Nakariakov, V.~M., \& Shibasaki, K.\ 2012, \apjl, 756, L36 
\bibitem[Luna-Cardozo et al.(2012)]{2012ApJ...748..110L} Luna-Cardozo, M., Verth, G., \& Erd{\'e}lyi, R.\ 2012, \apj, 748, 110
\bibitem[McIntosh et al.(2011)]{2011Natur.475..477M} McIntosh, S.~W., de Pontieu, B., Carlsson, M., et al.\ 2011, \nat, 475, 477
\bibitem[Macnamara \& Roberts(2010)]{2010A&A...515A..41M} Macnamara, C.~K., \& Roberts, B.\ 2010, \aap, 515, A41 
\bibitem[McEwan et al.(2006)]{2006A&A...460..893M} McEwan, M.~P., Donnelly, G.~R., D{\'{\i}}az, A.~J., \& Roberts, B.\ 2006, \aap, 460, 893 
\bibitem[McEwan et al.(2008)]{2008A&A...481..819M} McEwan, M.~P., D{\'{\i}}az, A.~J., \& Roberts, B.\ 2008, \aap, 481, 819
\bibitem[Mathioudakis et al.(2006)]{2006A&A...456..323M} Mathioudakis, M., Bloomfield, D.~S., Jess, D.~B., Dhillon, V.~S., \& Marsh, T.~R.\ 2006, \aap, 456, 323
\bibitem[Mathioudakis et al.(2003)]{2003A&A...403.1101M} Mathioudakis, M., Seiradakis, J.~H., Williams, D.~R., et al.\ 2003, \aap, 403, 1101
\bibitem[Mathioudakis et al.(2013)]{2013SSRv..175....1M} Mathioudakis, M., Jess, D.~B., \& Erd{\'e}lyi, R.\ 2013, \ssr, 175, 1 
\bibitem[Mendoza-Brice{\~n}o \& Erd{\'e}lyi(2006)]{2006ApJ...648..722M} Mendoza-Brice{\~n}o, C.~A., \& Erd{\'e}lyi, R.\ 2006, \apj, 648, 722 
\bibitem[Morton \& Erd{\'e}lyi(2009)]{2009ApJ...707..750M} Morton, R.~J., \& Erd{\'e}lyi, R.\ 2009, \apj, 707, 750
\bibitem[Mitra-Kraev et al.(2005)]{2005A&A...436.1041M} Mitra-Kraev, U., Harra, L.~K., Williams, D.~R., \& Kraev, E.\ 2005, \aap, 436, 1041 
\bibitem[Nakariakov et al.(1999)]{1999Sci...285..862N} Nakariakov, V.~M., Ofman, L., Deluca, E.~E., Roberts, B., \& Davila, J.~M.\ 1999, Science, 285, 862
\bibitem[Nakariakov et al.(2004)]{2004A&A...414L..25N} Nakariakov, V.~M., Tsiklauri, D., Kelly, A., Arber, T.~D., \& Aschwanden, M.~J.\ 2004, \aap, 414, L25
\bibitem[Nakariakov \& Verwichte(2005)]{2005LRSP....2....3N} Nakariakov, V.~M., \& Verwichte, E.\ 2005, Living Reviews in Solar Physics, 2, 3
\bibitem[Nakariakov \& Melnikov(2009)]{2009SSRv..149..119N} Nakariakov, V.~M., \& Melnikov, V.~F.\ 2009, \ssr, 149, 119 
\bibitem[Nakariakov \& Zimovets(2011)]{2011ApJ...730L..27N} Nakariakov, V.~M., \& Zimovets, I.~V.\ 2011, \apjl, 730, L27 
\bibitem[Ofman \& Wang(2002)]{2002ApJ...580L..85O} Ofman, L., \& Wang, T.\ 2002, \apjl, 580, L85 
\bibitem[O'Shea et al.(2001)]{2001A&A...368.1095O} O'Shea, E., Banerjee, D., Doyle, J.~G., Fleck, B., \& Murtagh, F.\ 2001, \aap, 368, 1095 
\bibitem[O'Shea et al.(2007)]{2007A&A...473L..13O} O'Shea, E., Srivastava, A.~K., Doyle, J.~G., \& Banerjee, D.\ 2007, \aap, 473, L13 
\bibitem[Pandey \& Srivastava(2009)]{2009ApJ...697L.153P} Pandey, J.~C., \& Srivastava, A.~K.\ 2009, \apjl, 697, L153
\bibitem[Pandey \& Singh (2009)]{2008MNRAS.387.1627P} Pandey, J.~C., \& Singh K. P. 2008, \mnras, 387, 1627
\bibitem[Reale (1997)]{1997A...325..782R}Reale, F., Betta, R., Peres, G., Serio, S., \& McTiernan, J.\ 1997, \aap, 325, 782
\bibitem[Reale(2007)]{2007A&A...471..271R} Reale, F.\ 2007, \aap, 471, 271 
\bibitem[Ruderman(2011)]{2011SoPh..271...41R} Ruderman, M.~S.\ 2011, \solphys, 271, 41 
\bibitem[Scargle(1982)]{1982ApJ...263..835S} Scargle, J.~D.\ 1982, \apj, 263, 835
\bibitem[Selwa et al.(2007)]{2007ApJ...668L..83S} Selwa, M., Ofman, L., \& Murawski, K.\ 2007, \apjl, 668, L83
\bibitem[Srivastava et al.(2008)]{2008MNRAS.388.1899S} Srivastava, A.~K., Zaqarashvili, T.~V., Uddin, W., Dwivedi, B.~N., \& Kumar, P.\ 2008, \mnras, 388, 1899
\bibitem[Srivastava \& Dwivedi(2010)]{2010NewA...15....8S} Srivastava, A.~K., \& Dwivedi, B.~N.\ 2010, \na, 15, 8 
\bibitem[Stepanov et al.(2001)]{2001A&A...374.1072S} Stepanov, A.~V., Kliem, B., Zaitsev, V.~V., et al.\ 2001, \aap, 374, 1072 
\bibitem[Taroyan \& Bradshaw(2008)]{2008A&A...481..247T} Taroyan, Y., \& Bradshaw, S.\ 2008, \aap, 481, 247 
\bibitem[Torrence \& Compo (1998)]{1998BAMS...79...61T} Torrence, C. \& Compo, G.P. \ 1998, BAMS, 79, 61
\bibitem[Verth \& Erd{\'e}lyi(2008)]{2008A&A...486.1015V} Verth, G., \& Erd{\'e}lyi, R.\ 2008, \aap, 486, 1015
\bibitem[Verwichte et al.(2009)]{2009ApJ...698..397V} Verwichte, E., Aschwanden, M.~J., Van Doorsselaere, T., Foullon, C., \& Nakariakov, V.~M.\ 2009, \apj, 698, 397 
\bibitem[Wang(2011)]{2011SSRv..158..397W} Wang, T.\ 2011, \ssr, 158, 397
\end{thebibliography}

\clearpage
\clearpage
\begin{deluxetable}{ccccc}
\tabletypesize{\normalsize}
\tablecaption{The summary of detected periodicities using wavelet analysis.}
\tablewidth{0pt}
\tablehead{
\colhead{Binning time} & \colhead{Cycles} & \colhead{Periods} & \colhead{Probability}
}
\startdata
50 s & $>$4 & 1274 s \& 694 s & $>$99\% \\
60 s & $>$4 & 1286 s \& 701 s & $>$99\% \\
70 s & $>$4 & 1261 s \& 687 s & $>$99\% \\
90 s & $>$4 & 1250 s \& 682 s & $>$99\% \\
\enddata
\end{deluxetable}

\begin{figure}
\centering
\includegraphics[scale=0.60, angle=0]{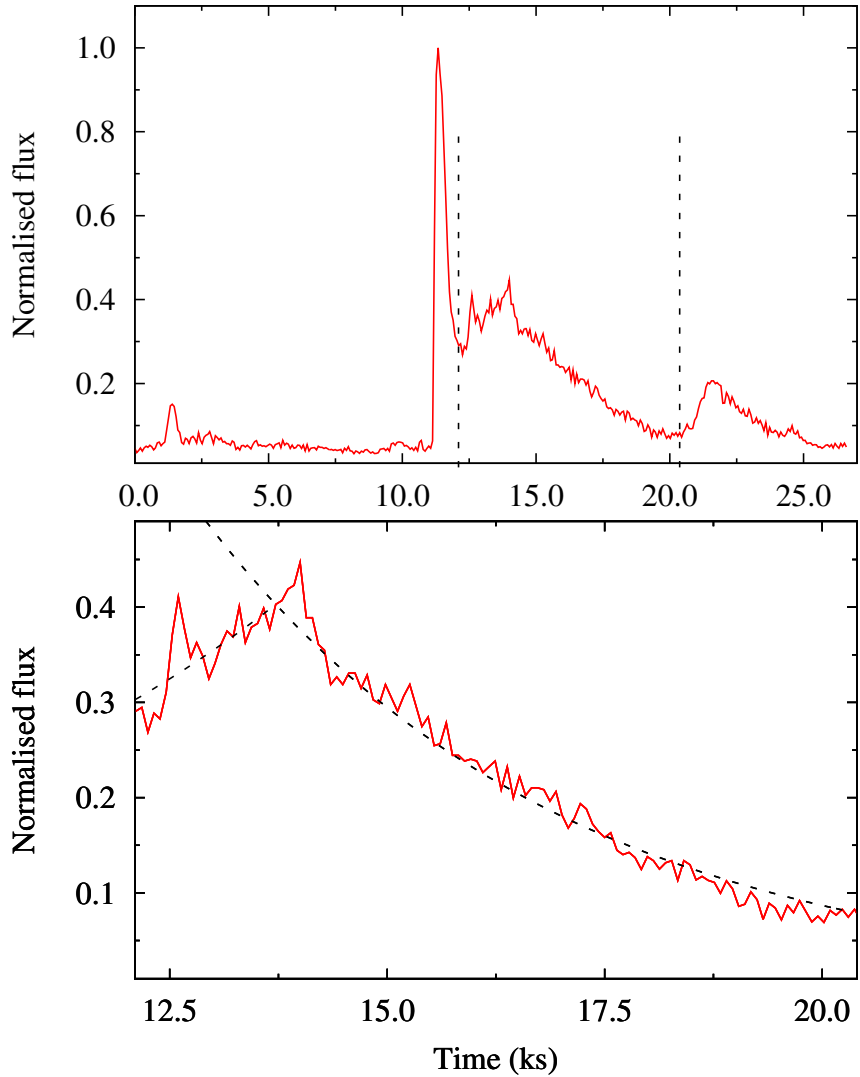}
\vspace{-0.50cm}
\caption{\small
Top-panel: XMM-Newton light curve of Proxima Centauri 
in 0.3-10~keV energy band with binning time of 70s. 
The detected oscillations are localized during 12-21 ks as shown
between two vertical dotted-lines.
Bottom panel: The best-fit exponential function along with the light curve under the dashed line as shown in top panel. The best fit exponentials were used to remove long-term trends associated with the flare background effects.
}
\label{fig:JET-PULSE}
\end{figure}
%
\clearpage
\clearpage
\begin{figure}
\centering
\includegraphics[scale=0.75, angle=90]{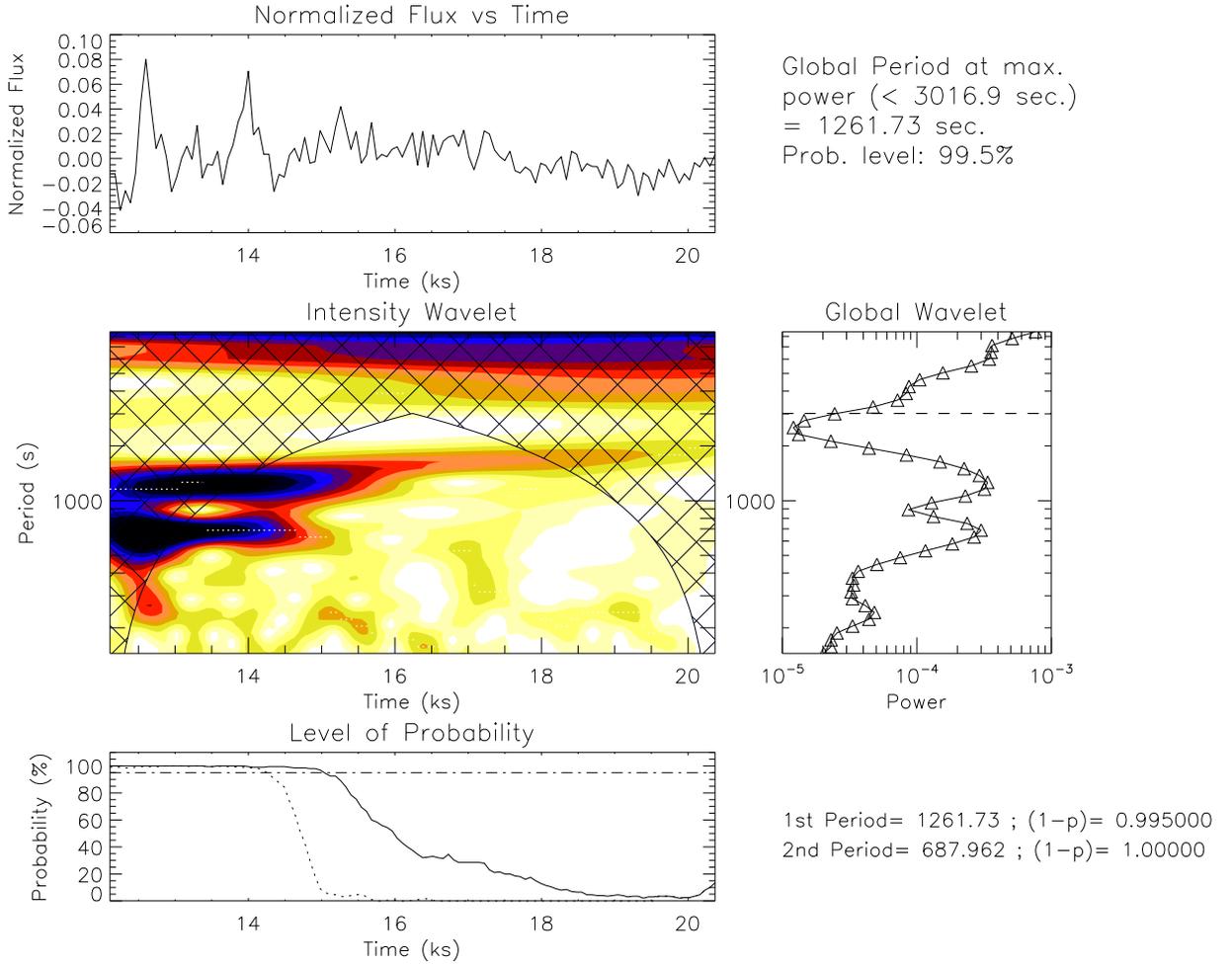}
\caption{\small  The wavelet result :Top panel shows the 
detrended light curve removed from flare background; Middle-left panel shows the intensity wavelet;
Middle-right panel shows the global power spectrum with a significant 
periodicity of 1261 s with the probability of 99.5 \%.
Other period and related power is concentrated around $\sim$
687 s in the intensity wavelet.
The bottom panel shows the probability variation of the two detected periodicities.} 
\label{fig:JET-PULSE}
\end{figure}

\clearpage
\begin{figure}
\centering
\hspace{4.8cm}
\includegraphics[scale=0.6, angle=-90]{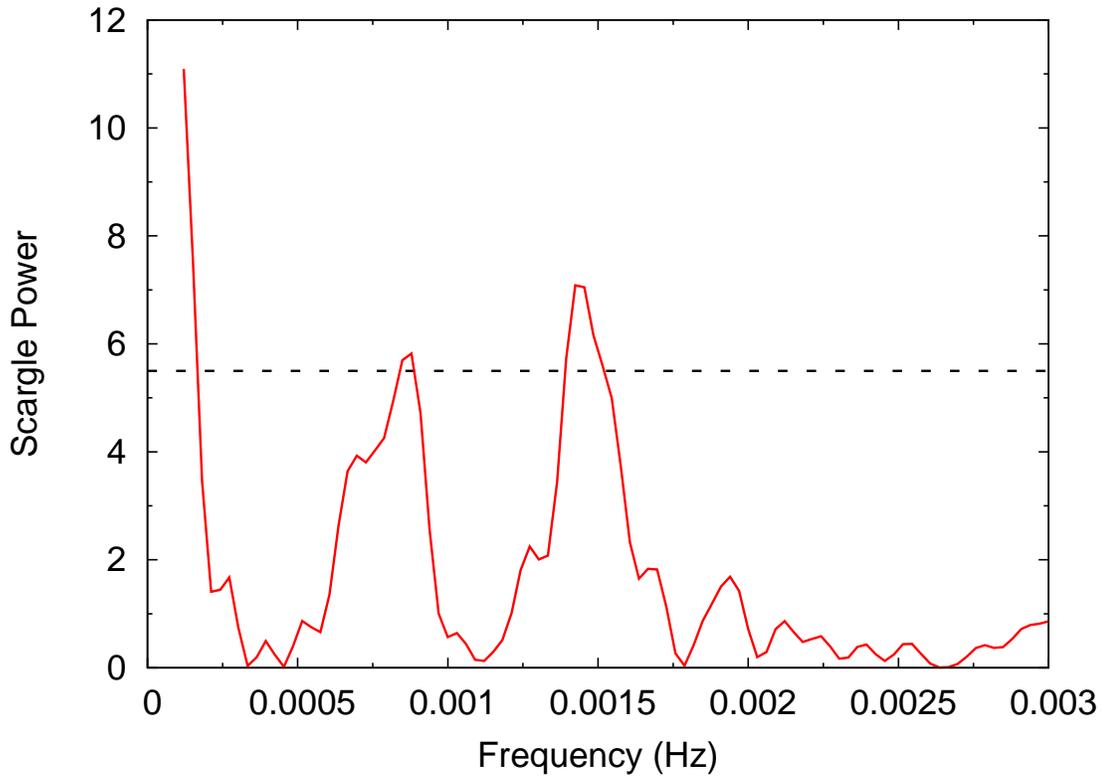}
\caption{\small
Power spectra of detrended light curve  as shown in top panel of Fig. \ref{fig:JET-PULSE}.
The horizontal line is 1-$\sigma$ level significance. The detection of edged and
unrealistic period of 8000 s is also evident in the periodogram that suppresses the power and significance of two naturally evolved periods.} 
\label{fig:JET}
\end{figure}
%
\begin{figure}
\centering
\includegraphics[scale=0.50, angle=-90]{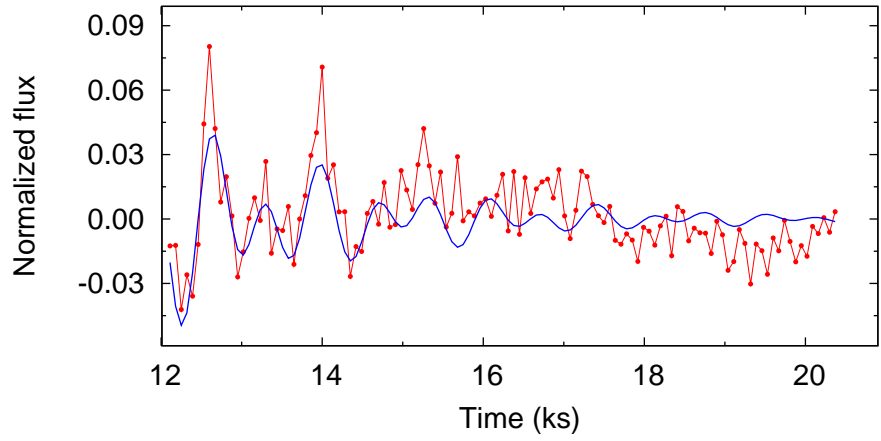}
\caption{\small
The detrended light curve of the post-flare phase along with
the best fit exponentially decaying harmonic function as given by Eq.~1.
}
\label{fig:PULSE}
\end{figure}

%

\end{document}